\documentclass[
twocolumn,
reprint,
superscriptaddress,
nofootinbib,
nobibnotes,
amsmath,
amssymb,
aps,
pra,
longbibliography,
]{revtex4-2}
 
\usepackage[latin1]{inputenc}
\usepackage{amsmath}
\usepackage{amsfonts}
\usepackage{hyperref}
\usepackage{amssymb}
\usepackage{graphicx}
\usepackage{hyperref}
\usepackage{colortbl}
\usepackage{subfigure}
\usepackage{svg}

\usepackage[normalem]{ulem}

\usepackage[T1]{fontenc}
\usepackage{mathptmx}
\usepackage{bbold}
\usepackage{scalerel}
\usepackage{xcolor}

\usepackage{mathtools}

\usepackage{chemformula}
\newcommand{\hyref}[1]{\hyperref[#1]{\ref{#1}}}

\newcommand{\orange}[1]

\usepackage{titlesec}

\renewcommand{\thesection}{\arabic{section}}

\titleformat{\section}  % which section command to format
  {\fontsize{10}{12}\bfseries} % format for whole line
  {\thesection} % how to show number
  {1em} % space between number and text
  {\centering} % formatting for just the text
  [] % formatting for after the text

\titleformat{\subsection} % which command to format
  {\fontsize{9}{11}\bfseries} % font size and style
  {\thesubsection}            % subsection numbering
  {1em}                       % space between number and text
  {}                          % formatting applied only to the title text (empty = left aligned)
  []             

\begin{document}

\title{Optical Voltammetry of redox processes inside a nanohole with opto-iontronic microscopy}

\author{Zhu Zhang}
\thanks{These two authors contributed equally to this work}
% \thanks{Corresponding author}
\affiliation{Nanophotonics, Debye Institute for Nanomaterials Science, Department of Physics, Utrecht University, Princetonplein 1, 3584 CC, Utrecht, The Netherlands}
 
\author{\normalfont\textsuperscript{,$\dag$}\;Haolan Tao}
\thanks{These two authors contributed equally to this work}
\affiliation{State Key Laboratory of Chemical Engineering, School of Chemical Engineering, East China University of Science and Technology, 130 Meilong Road, Shanghai, 200237, China}
\author{Cheng Lian}
\affiliation{State Key Laboratory of Chemical Engineering, School of Chemical Engineering, East China University of Science and Technology, 130 Meilong Road, Shanghai, 200237, China}
\author{Ren\'e van Roij}
\thanks{Corresponding author}
\affiliation{Institute for Theoretical Physics, Department of Physics, Utrecht University, Princetonplein 5, 3584 CC, Utrecht, The Netherlands}
\author{Sanli Faez}
\thanks{Corresponding author}
\affiliation{Nanophotonics, Debye Institute for Nanomaterials Science, Department of Physics, Utrecht University, Princetonplein 1, 3584 CC, Utrecht, The Netherlands}

\date{\today}
\begin{abstract}
Cyclic Voltammetry (CV) is the most commonly used method in electrochemistry to characterize electrochemical reactions, usually involving macroscopic electrodes.
Here we demonstrate an optical CV technique called Opto-iontronic  Microscopy, which is capable of monitoring electrochemical processes at the nanoscale. By integrating optical microscopy with nanohole electrodes, we enhance sensitivity in detecting redox reactions within volumes as small as an attoliter ($(100 \text{~nm})^{3}$). This technique uses electric-double-layer modulation and lock-in detection to sensitively probe ion dynamics during cyclic voltammetry in nanoholes that are under total internal reflection illumination. We applied this method to study electric double layer (dis)charging coupled to ferrocenedimethanol (\ch{Fc(MeOH)2}) redox reactions. Experimental results were validated against a theoretical Poisson-Nernst-Planck-Butler-Volmer model, providing insights into ion concentration changes of reaction species that contribute to the optical contrast. This work opens up opportunities for high-sensitivity, label-free analysis of electrochemical reactions in nanoconfined environments, with potential applications in pure nanocrystal growth and monitoring.
\end{abstract}

\maketitle

The electric double layer (EDL) formed at charged solid-liquid interfaces significantly influences electrochemical reactions occurring at a solid electrode in contact with a liquid electrolyte. 
This influence becomes even more pronounced in the case of confinement on the nm-scale, where the ions within the EDL and the ion transport processes become closely coupled. This coupling can dramatically alter the behavior and efficiency of electrochemical conversion at the nanoscale compared to macroscopic electrodes.

Two of the most commonly used methods for characterizing electrochemical processes at an interface are Cyclic Voltammetry (CV)~\cite{wangPhysicalInterpretationCyclic2012} and Electrochemical Impedance Spectroscopy  (EIS)~\cite{wangElectrochemicalImpedanceSpectroscopy2021}. 
In CV one applies a slowly varying electric potential to a metal electrode, which is in contact with the electrolyte solution, and monitors the resulting currents in the external circuit.
In EIS, one applies a low-amplitude sinusoidal potential to the electrodes over a range of frequencies and calculates the frequency-dependent (complex) impedance of the chemical cell based on the measured amplitude and phase of the alternating electric current passing through the external circuit~\cite{wangElectrochemicalImpedanceSpectroscopy2021}.

%- Votammetry is a major analytical technique in chemistry
Other surface characterization methods such as the electrochemical quartz crystal microbalance (EQCM) and ellipsometry have been employed as very powerful \textit{in situ} techniques to complement electrochemical experiments~\cite{leviApplicationQuartzcrystalMicrobalance2009, freundElectrochemicalQuartzCrystal1990, barisciElectrochemicalQuartzCrystal2000, minenkovMonitoringElectrochemicalFailure2024, brochRealTimeSitu2007} to monitor EDL charging dynamics by detecting mass changes at a quartz resonator electrode~\cite{parsonsBandStructureAssociated1969, nomuraFrequencyShiftsPiezoelectric1982, EffectElectricalDouble1968}.
Although these methods can indeed monitor characteristics of the EDL structure very precisely and can achieve a high temporal resolution, they cannot, however, spatially resolve the local electrochemical properties such as heterogeneities of the electrode surface, which majorly complicates the interpretation of the measured results.
Local electrochemical properties can be probed by the scanning electrochemical microscope (SECM) as invented by Bard \textit{et al.}~\cite{bardChemicalImagingSurfaces1991} to image the sample topography and surface reactivity. 
Later on SECM was used to measure surface charge  densities~\cite{klausenMappingSurfaceCharge2016, mckelveySurfaceChargeMapping2014b} and EDL dynamics~\cite{hurthDirectProbingElectrical2007,tanDoubleLayerEffects2018} as well as redox reactions~\cite{duNanoscaleRedoxMapping2021}. 
However, because SECM is very sensitive to the distance between the probe and the surface, it is a challenging method for measurement on a relatively large area of, say,  10~$\mu$m$^2$ ~\cite{wittstockScanningElectrochemicalMicroscopy2007}.
%To monitor electrochemical reactions , %many techniques have been invented.
Raman spectroscopy, which monitors the spectrum of the scattered light from catalytic processes, has also been widely used for probing electro-catalysis at solid-liquid interfaces. 
For instance, 
Wang \textit{et al.} used enhanced Raman spectroscopy to reveal the structure and dissociation of interfacial water~\cite{wangSituRamanSpectroscopy2021a}, and
Yin \textit{et al.} reported tip-enhanced Raman spectroscopy (TERS) to study the catalytic hydrogenation of chloronitrobenzenethiol on a Pd/Au bimetallic catalyst~\cite{yinNanometrescaleSpectroscopicVisualization2020a}.
However, the measurement speed of Raman spectroscopy depends on the integration time necessary for collecting enough light to determine the spectrum, and getting a high-SNR (Signal-Noise Ratio) spectrum usually comes at the expense of lowering the time resolution.
%- The potentiodynamic optical contrast has been used in various configuration for obtaining surface electrochemical information, although so far with optical resolution.

%- advantage of optical microscope is parallel imaging, compatible with surface dynamics and other optical investigation modalities.
In recent years, optical microscopy has received renewed interest for probing electrochemical reactions at interfaces. 
Optical microscopy's intrinsic advantages of parallel imaging enable the probing of fast electrochemical reactions at high speed with high spatial resolution~\cite{ElectrochemicalIonInsertionReactions, evans_influence_2019, lemineur_situ_2020}.
Recently, for instance, Evans \textit{et al.} imaged single-nanoparticle electrochromic dynamics~\cite{evans_influence_2019} and Moghaddam \textit{et al.} probed the local paths of charge transfer of copper hexacyanoferrate (CuHCF) microparticles~\cite{moghaddam_scanning_2024} by optical absorption microscopy. Hu \textit{et al.} used dark-field scattering microscopy to observe atomic layer electro-deposition on a single nanocrystal surface~\cite{huObservingAtomicLayer2020a}, and Altenburger \textit{et al.} used it for catalytic \ch{H2O2} decomposition imaging~\cite{altenburger_label-free_2023}.
Wang \textit{et al.} have directly imaged dynamic electronic coupling during the electrochemical oxidation of single silver nanoparticles~\cite{jiangDirectlyImagingDynamic2022}. 
They also monitored the surface charging depth of single Prussian blue nanoparticles from their scattering light intensities by a modulation potential~\cite{niuDeterminingDepthSurface2022}.
Earlier, optical reflection was implemented to image the electrochemical activities on the electrode surface by using interferometric methods~\cite{liInterferometricMeasurementDepletion1995a, flatgenTwoDimensionalImagingPotential1995, anderssonImagingSPRDetection2008} or using surface plasmon imaging~\cite{shanImagingLocalElectrochemical2010}.
Plasmonic-enhanced measurements, however, are mainly suitable for metallic materials that exhibit a plasmonic resonance~\cite{kirchnerSnapshotHyperspectralImaging2018, byersSingleParticleSpectroscopyReveals2014,hoenerSpectroelectrochemistryHalideAnion2016, byersTunableCoreshellNanoparticles2015, hoenerPlasmonicSensingControl2018}. 

Recently, our group demonstrated a new technique called EDL modulation microscopy, which is based on the time-dependent optical contrast generated by a periodic modulation of the potential of an electrode that is in contact with an electrolyte~\cite{naminkElectricDoubleLayerModulationMicroscopy2020}. In this method, modulating the EDL close to the charged surface results in a differential scattering signal that is sensitive not only to the local topography but also to the electrochemical properties of the investigated region~\cite{zhangComputingLocalIon2021}. 
Unlike previous spectro-electrochemical investigations of nanoparticles, EDL-modulation microscopy can be applied to non-plasmonic, dielectric particles and biological samples as well~\cite{kurosu_spatiotemporal_2024}.

%- we shrink the imaging volume to sub-wavelength nanoholes
%- this is not trivial because signal scales with volume, but interferometric scattering allows for it

%- we recover the electrochemical activity of the ferrocene
%- we compare experimental results with theory
%- concentration of active species is the main indicator of the contrast

In this paper, we present further development of this microscopy method to monitor electrochemical activity in aqueous electrolytes inside nanoholes. 
We classify this method under the broader category of ``opto-iontronic microscopy'' and choose the standard and well-characterized redox species 1,1-Ferrocenedimethanol (\ch{Fc(MeOH)2}), dissolved in 0.1~M~\ch{KCl}, for our measurements.
By using sub-optical-wavelength nanoholes with height and diameter of about~$100$~nm, we shrink the imaging volume of the electrochemical reactions down to the attoliter ($(100 \text{~nm})^{3}$) scale. 
To get a decent signal from such an extremely small volume is not trivial, because the optical scattering signal is proportional to the volume of the imaged object. 
Therefore we combine the EDL-modulation technique with optical lock-in detection to enhance the signal-to-noise ratio.
First, as will be explained in full detail in the following, we performed EDL charging measurements inside a single nanohole, fabricated by a focused ion beam (FIB), and
investigated the potentiodynamic contrast by slowly scanning the potential applied to the nanohole electrode from $-0.2$~V to $0.2$~V in absence of the redox species.
Next, we investigated the resulting light scattering intensities in the presence of the electrochemical reactions inside the nanohole with the lock-in detection technique. 
We also present a theoretical model to calculate the time-dependent ion concentration profiles inside the nanohole during the electrochemical reactions. 
Comparing the theoretical predictions with the experimental results strongly suggests that the concentration of the active redox species inside the nanohole gives the main contribution to the optical contrast. 
In other words: we have an optical detection of the electrochemical process on the attoliter scale.

\section{Opto-iontronic microscopy development}
\begin{figure*}[htbp]
\centering
\includegraphics[width=\textwidth]{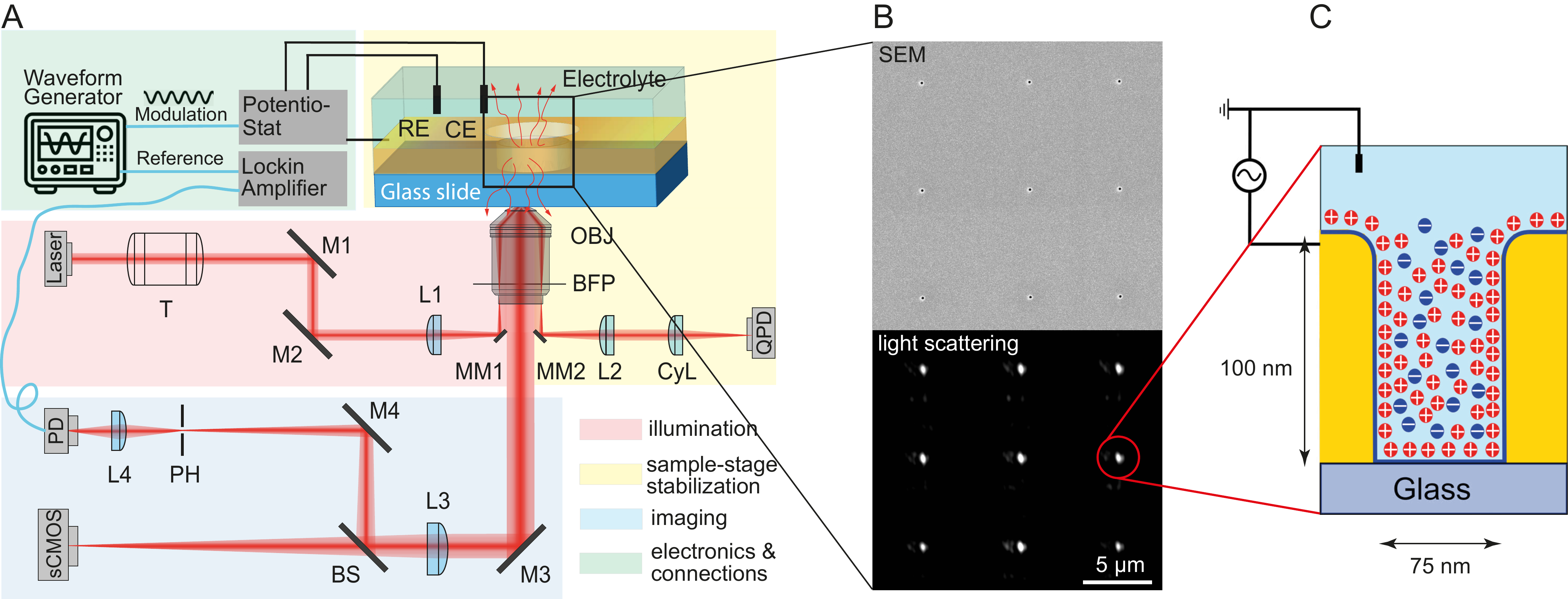}
\caption{ (A) Schematic of the total\--internal\--reflection opto\--iontronic microscope. T: telescope for beam diameter adjustment; M1\--M4: adjustable mirrors; L1, L2: beam focus lens; L3, L4: imaging lens; MM1, MM2: small prism mirrors; OBJ: microscope objective; BFP: back focal plane of the microscope objective; CyL: cylindrical lens; RE: reference electrode in the chemical cell; CE: counter electrode in the chemical cell; BS: beam splitter; PH: pinhole; PD: photodiode; sCMOS: scientific complementary metal\--oxide\--semiconductor camera; QPD: quadrant photodiode. (B) SEM image (top) and light scattering image (bottom) of the nanohole array. The light scattering image was taken by the sCMOS camera without an applied potential. (C) Schematic representation of the nanohole geometry, with a depth of $100$~nm and a diameter of $75$~nm.  }\label{fig_Setup_SEM_CMOS}
\end{figure*}

Our opto-iontronic microscopy of the electrode-electrolyte interface is based on the evanescent light field that is generated from total internal reflection (TIR) resulting from the illumination of a glass slide that supports an electrode with an array of nanoholes, as shown in Fig.~\ref{fig_Setup_SEM_CMOS}A. 
%Our iontronic microscopy method is based on a total-internal-reflection (TIR) illumination regime. The evanescent light field generated from TIR illuminates objects close to the interface between the glass slide and the liquid solution, shown in Fig.~\ref{fig_Setup_SEM_CMOS}(a).
This method is similar to Total-Internal-Reflection-Fluorescence (TIRF) microscopy, which excites a fluorescent object at the interface with an evanescent field and images its corresponding fluorescence emission light~\cite{axelrodTotalInternalReflection2001}.
Rather than imaging the fluorescence emission light as in TIRF microscopy, our experimental instrument collects the scattered light. 
Our instrument thus allows us to image any object close to the interface, without the need for it to be fluorescent. 
Meng \textit{et al.} implemented this method to track single gold nanoparticles at oil-water interfaces~\cite{mengMicromirrorTotalInternal2021} and Namink \textit{et al.} used it to image EDL charging/discharging dynamics around small ITO nanoparticles during potential modulation~\cite{naminkElectricDoubleLayerModulationMicroscopy2020, petersDarkfieldLightScattering2023}.
We improved the detection sensitivity by several orders of magnitude by the implementation of a photodiode and a lock-in amplifier in the experimental instrument, which enhances the detection sensitivity from ~$10^{-3}$~\cite{naminkElectricDoubleLayerModulationMicroscopy2020, mengMicromirrorTotalInternal2021} to ~$10^{-6}$, where sensitivity is defined as the ratio of the optical modulation amplitude to the baseline scattered light intensity.

The electrode with an array of nanoholes was fabricated by drilling into a $100$~nm Au layer (supported by the glass slide) using a focused ion beam (FIB), see details in the Methods section. Fig.~\ref{fig_Setup_SEM_CMOS}B presents a top view SEM image of a part of the array of nanoholes alongside the light scattering image of the nanoholes, both taken without an applied potential. The nanoholes form a square lattice with a lattice spacing of about 6 micron, and the essentially cylindrical geometry of the individual nanoholes is schematically illustrated in the insets of Fig.~\ref{fig_Setup_SEM_CMOS}C, with a depth of $100$~nm and a diameter of $75$~nm, with deviations of the order of 10\%.
Compared with a planar electrode, the nanohole electrode offers a significantly higher surface-to-volume ratio, which enhances the contribution of ions from the EDL region.
Because the evanescent wave at the glass\--gold interface only penetrates to a depth of about $100$~nm into the gold, it effectively illuminates the entire nanohole while blocking unwanted optical signals from the gold\--water interface. Moreover, the sub--diffraction--limited size of the nanoholes confines the optical signal to the space within each nanohole, restricting the measurement volume and improving the detection sensitivity. 

\section{EDL charging/discharging  and electrochemical redox reactions}

\begin{figure*}[ht]
\centering
\includegraphics[width=\textwidth]{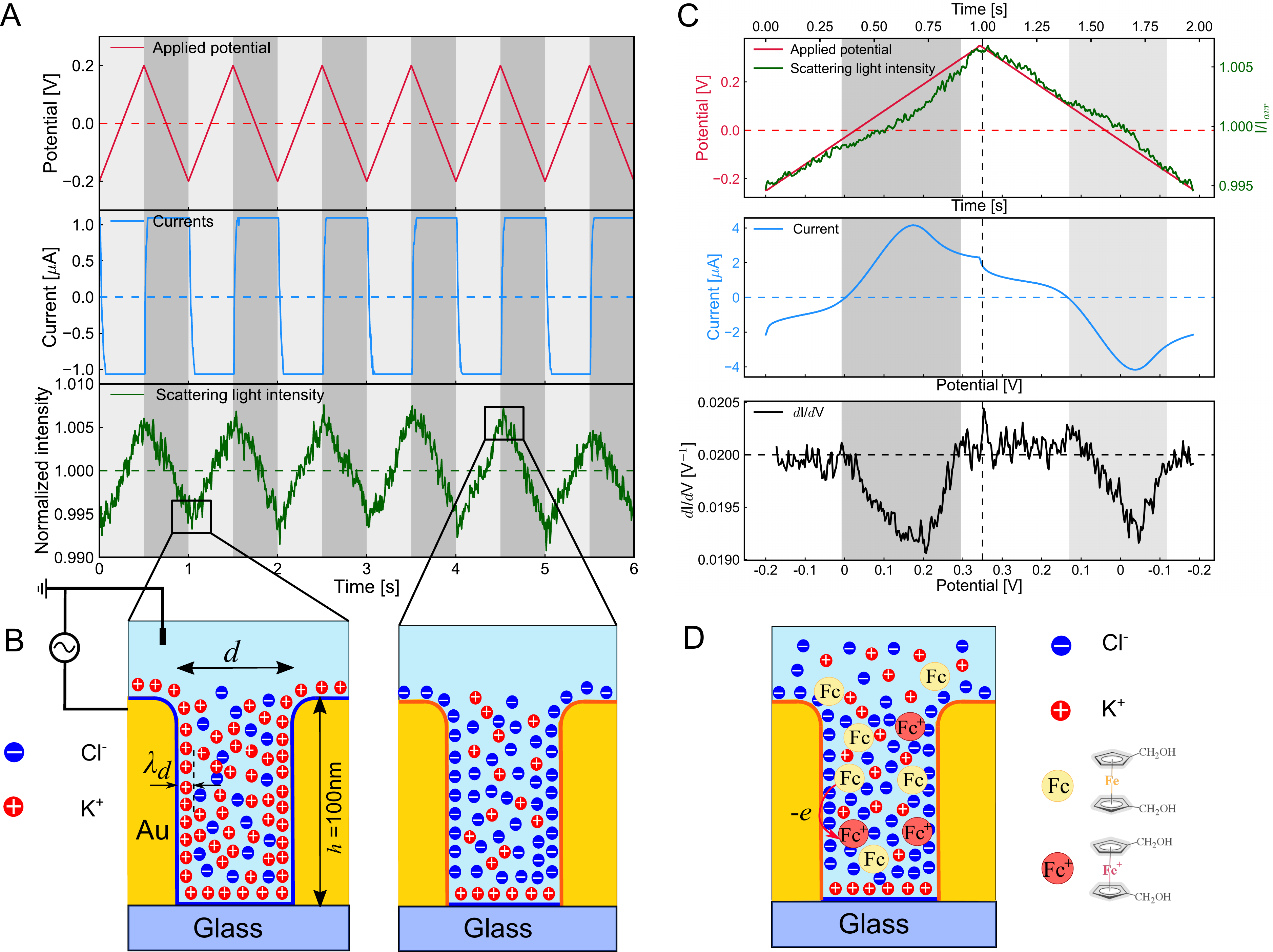}
\caption{Potentiodynamic measurements of EDL charging/discharging in parallel with cyclic voltammetry inside a single nanohole. (A) Capacitive EDL charging/discharging of a single nanohole in contact with a grounded bulk~\ch{KCl} solution of concentration $C_{\mathrm{KCl}} = 0.1\text{~M}$. The potential waveform $V(t)$ (top panel) is applied to the nanohole electrode (yellow in the insets),  the corresponding capacitive current $i(t)$ (middle panel) from the nanohole electrode and the intensity $I(t)$ (bottom) of the light scattered from a single nanohole normalized to the intensity at potential $V=0$~V. (B) Sketches of the ion arrangements at negative potential (left) and positive potential (right) on the inner surface of the nanohole, featuring the EDL of thickness $\lambda_d\simeq 1\text{~nm}$ (not to scale) that forms the capacitor. (C) Cyclic voltammetry measurement inside the nanohole with 1 mM~\ch{Fc(MeOH)2} and 0.1~M~\ch{KCl} supporting electrolyte, where the potential is scanned back and forth between $-0.25$~V and  $0.35$~V with a scanning speed of 600~mV/s. Top panel: The scan potential $V(t)$ (red) applied to the nanohole electrode and the corresponding normalized light scattering intensity $I(t)$ of the nanohole averaged over 100 cycles (green). Middle panel:  The potential dependence of the total current $i(V(t))$, which changes sign upon entering the oxidation (dark gray) and reduction (light gray) windows due to Faradaic contributions that develop on top of the capacitive current. Bottom panel: The variation of the light scattering intensity with applied electrode potential, $dI/dV$, which drops below the constant capacitive baseline (dashed horizontal) in the two redox windows, signifying the optical measurement of the electrochemical redox processes in the nanohole.  (D) Sketch of the ion and redox species arrangements in the nanohole in the oxidation window (positive potential).}\label{fig_DC_CV}
\end{figure*}

We investigated the potentiodynamic contrast by scanning the potential applied to the electrode back and forth between $-0.2$~V to $0.2$~V with a scan rate of $800$~mV/s, \textit{i.e.} the frequency of the applied periodic potential modulation is as low as $1$ Hz. The electrode is in contact with a grounded bulk~\ch{KCl} solution of bulk concentration $C_{\mathrm{KCl}} = 0.1$~M.
For the case without any redox reaction, i.e. without \ch{Fc(MeOH)2}, Fig.~\ref{fig_DC_CV}A shows the time-dependence of the triangular waveform $V$ of the applied potential (red), the resulting current $i$ in the external circuit (blue), and the (normalized) intensity $I$ of the light scattered from a single nanohole (green). We note that a constant current $i=\pm 1~\mu\text{A}$ to/from the total electrode with surface area $A_{\mathrm{tot}} = 2.5\,\mathrm{mm}^2$ at a constant scan rate $dV/dt=\pm 800$~mV/s is consistent with the (Ohmic, non-Faradaic) charging/discharging of a capacitor of fixed capacitance $C=i/(dV/dt)=1.25~\mu\text{F}$. This capacitance is many orders of magnitude larger than to be expected from the EDL that covers a single nanohole, since the current is collected from the total electrode. The (normalised) optical intensity $I$, however, stems from only a single nanohole and is seen to satisfy $dI\propto dV$ in this purely capacitive case. 

As schematically illustrated in Fig.~\ref{fig_DC_CV}B, the cations (\ch{K+}) accumulate on the inner surface of the nanohole when the electrode potential is negative. As the potential is swept to positive, the anions (\ch{Cl-}) gradually replace the cations to screen the surface.
Due to the difference in the optical polarizability of the dissolved~\ch{K+} and~\ch{Cl-} ions, the ion concentration profile inside the nanohole during the cycle of the potential will also change the effective refractive index of the whole volume of the nanohole. Thus, by periodically changing the electrode potential, we also obtain the observed periodic change in the intensity of the scattered light, which is in fact in agreement with the potentiodynamic contrast measurement of ITO grain nanoparticles as previously presented by Namink \text{et al.}~\cite{naminkElectricDoubleLayerModulationMicroscopy2020}. 
%Note the small variation of less than a percent of the light intensity, which is nevertheless sufficiently large for detection by our sensitivity-enhanced device.

After having established that our optical measurements can be directly related to the ionic composition inside the nanohole, we investigated the light scattering intensities in response to electrochemical reactions inside the nanoholes. We again bring the electrode in contact with the same grounded $0.1$~M~\ch{KCl} solution, however now with $1$~mM~\ch{Fc(MeOH)2} added that can undergo the standard redox reaction ~\ch{Fc(MeOH)2}~$\underset{+e}{\stackrel{-e}{\rightleftharpoons}}$~\ch{Fc(MeOH)2+} in the vicinity of the electrode.
%To make a chemical reaction happen inside the nanohole, we chose a standard well-characterized redox species, Ferrocene-dimethanol (\ch{Fc(MeOH)2}) with reaction~\ch{Fc(MeOH)2}~$\underset{-e}{\stackrel{+e}{\rightleftharpoons}}$~\ch{Fc(MeOH)2+}, for our measurements.
Fig.~\ref{fig_DC_CV}C shows the light scattering intensity profile (green), averaged over 100 cycles of 2~s periods, as a function of the electrode potential (red) during cyclic voltammetry (CV) with a triangular potential that varies back and forth between $-0.25$~V and $+0.35$~V with a frequency of 0.5~Hz. 
The blue curve of the middle panel of Fig.~\ref{fig_DC_CV}C shows the corresponding current $i$ in the external circuit, also as a function of the rising and lowering potential. Rather than a constant positive (negative) current at lowering (rising) potential that switches rather abruptly at the extrema of the potential, as observed in (A) in the absence of any redox reaction, the current $i$ (blue) in panel Fig.~\ref{fig_DC_CV}C changes sign smoothly somewhere halfway the rise and fall of the potential. This is due to the redox reactions: during the rise of the potential $i$ becomes positive due to the appearance of an oxidation current that exceeds the (negative) EDL charging current, and during the fall of the potential $i$ becomes negative due to the appearance of a reduction current that exceeds the (positive) EDL discharging current. Thus, the purely capacitive relation $i\propto dV/dt$ breaks down in the presence of a redox reaction. During the forward potential scan, as the potential increases from negative to positive, it eventually reaches the oxidation potential of the oxidation reaction for \ch{Fc(MeOH)2}. At this point, \ch{Fc(MeOH)2} near the electrode surface is rapidly oxidized to \ch{Fc(MeOH)2+}, resulting in a sharp increase in current, observed as the oxidation peak in panel C. This rapid reaction depletes the local concentration of \ch{Fc(MeOH)2} and generates \ch{Fc(MeOH)2+} in the vicinity of the electrode.
As the scan continues, \ch{Fc(MeOH)2} from the bulk solution diffuses toward the electrode to sustain the oxidation reaction. This diffusion process limits the rate of reaction, leading to a plateau in the current that corresponds to the diffusion-limited (or mass transport limited) regime of the oxidation process.
During the reverse scan, as the potential is swept back toward negative values, it eventually reaches the reduction potential of \ch{Fc(MeOH)2+}. The oxidized ions (\ch{Fc(MeOH)2+}) near the electrode surface are therefore rapidly reduced back to \ch{Fc(MeOH)2}, producing a prominent reduction peak in panel C. As the potential continues to decrease, \ch{Fc(MeOH)2+} ions from the bulk diffuse back to the electrode surface to undergo reduction, leading to a second plateau in the current which is associated with the diffusion-limited reduction process.
Interestingly, simultaneously also the proportionality between the changes in voltage and the light intensity, $dI\propto dV$ as observed in panel A without redox reactions, breaks down in the presence of redox reactions, as can be seen by the deviation of the green curve ($I$) from the red curve ($V$) in the top panel of Fig.~\ref{fig_DC_CV}C, where $I$ is systematically decreased in the oxidation regime (dark grey) and systematically enhanced in the reduction regime (light grey). Outside these redox-dominated regimes the linearity $dI\propto dV$ is essentially retained.  
%which show significant deviation and the redox-driven deviation from the purely capacitive relation $i\propto dV/dt$ is also reflected by a deviation from also visible in the optical signal, which 
We can thus distinguish potential regimes in which the change of light scattering intensity scales essentially linearly with the potential change and regimes with a significantly decreased (dark grey) and enhanced (light grey) light intensity. In the linear regimes the current is also essentially linear with the potential, 
%between $-0.25$~V to 0.05~V,  
which is therefore an Ohmic (non-Faradaic) regime.
%The uptrend of the signal slows down in the oxidation window of~\ch{Fc(MeOH)2} (as shown in the low grey window) where the current increases dramatically due to the oxidation current of of~\ch{Fc(MeOH)2}. 
While the potential sweeps back to $-0.25$~V, the optical signal decreases and then the downtrend slows down in the reduction potential window (the high grey window in Fig.~\ref{fig_DC_CV}C) where the current decreases dramatically due to the reduction current of~\ch{Fc(MeOH)2}.
As the potential scans over the reduction window, the signal decreases linearly again.
The bottom panel of Fig.\ref{fig_DC_CV}C shows the derivative $dI/dV$ of the scattering light intensity with respect to the applied potential. The dashed line represents the capacitive baseline of the normalized intensity, derived from Fig.\ref{fig_DC_CV}A, where only EDL charging/discharging occurs. This baseline is used to define the light gray and dark gray regions in Fig.\ref{fig_DC_CV}C. When the derivative of the scattering intensity in Fig.\ref{fig_DC_CV}C drops below this baseline, the corresponding potential is considered to lie within the redox potential window.
We can see from the current signal that light scattering intensities are directly influenced by the supporting electrolyte ions charging/discharging on the inner surface of the nanohole and also by the redox reaction of~\ch{Fc(MeOH)2} inside the nanohole. 
Compared to the EDL charging/discharging cycle inside the nanohole with only~\ch{KCl} solution of Fig.~\ref{fig_DC_CV}B, the presence of the additional redox species inside the nanohole lead to a more complicated change of the local ion concentrations during the cycle, and hence to the more complicated optical response shown in Fig.~\ref{fig_DC_CV}C that we also illustrate schematically in Fig.~\ref{fig_DC_CV}D.  
On the basis of theoretical model predictions, presented below, we will show that the redox-active species are mostly responsible for the change of the optical response.

The results of Fig.\ref{fig_DC_CV}C are the first indication that information on the electrochemical reactions inside nanoholes can be extracted from charging/discharging cycles via their influence on the optical contrast. This optical effect can thus be used for monitoring electrochemical states and processes.
However, since the sCMOS camera that we used in our experiments has a high pixel density and a high signal-to-noise ratio, it is unfortunately also slow. In fact, with its achievable speed of only 200 frames per second (fps), it is a challenge to monitor the two key processes (EDL formation and electrochemical reactions) in real time without averaging over many cycles. In order to get a decent signal with this camera, we therefore usually take long measurements of 100 - 200 cycles and then average the optical signal. This averaging method allows us to get a good signal that can distinguish the electrochemical reactions from the EDL charging/discharging process, as shown in Fig.\ref{fig_DC_CV}, however, the long measurement duration (up to 30 minutes for each scan) is only suitable for very stable electrochemical conditions. 

In order to improve the sensitivity and the measurement speed of this opto-iontronic microscopy technique, we split the imaging arm and add a pinhole and a photodiode (PD) as shown in Fig.~\ref{fig_Setup_SEM_CMOS}A.
Not unlike confocal microscopy, the pinhole allows us to select a single nanohole, the scattered light of which is detected by a single photodiode that is in the image plane of the singlet lens behind the pinhole.
This photodiode has a higher bandwidth (2 kHz in this case) than the sCMOS camera and can also provide photon-to-current conversion at ultra-low noise levels. The time-dependent current variations from the converted light scattering intensities caused by sinusoidal modulations of the applied potential, as discussed below, are sent to a lock-in amplifier, from which the amplitude of the signal is retained. 

\section{Monitoring electrochemical redox reactions with opto-iontronic microscopy}
% \begin{figure}[htbp]
% \centering\includegraphics[width=\columnwidth]{Figures/EDL_Amp.VS.Freq.pdf}
% \caption{\label{fig_EDL__freqAmp}(a) Response of the amplitude of the optical signal in the function of the frequency of modulation potential. The amplitude of the modulation is $A_{Modulation} = 100~mV$. (b) Response of the amplitude of the optical signal to the amplitude of the modulation potential.The frequency of the modulation is $f_{Modulation} = 275~Hz$. }
% \end{figure}
\begin{figure*}[htbp]
\centering\includegraphics[width=\textwidth]{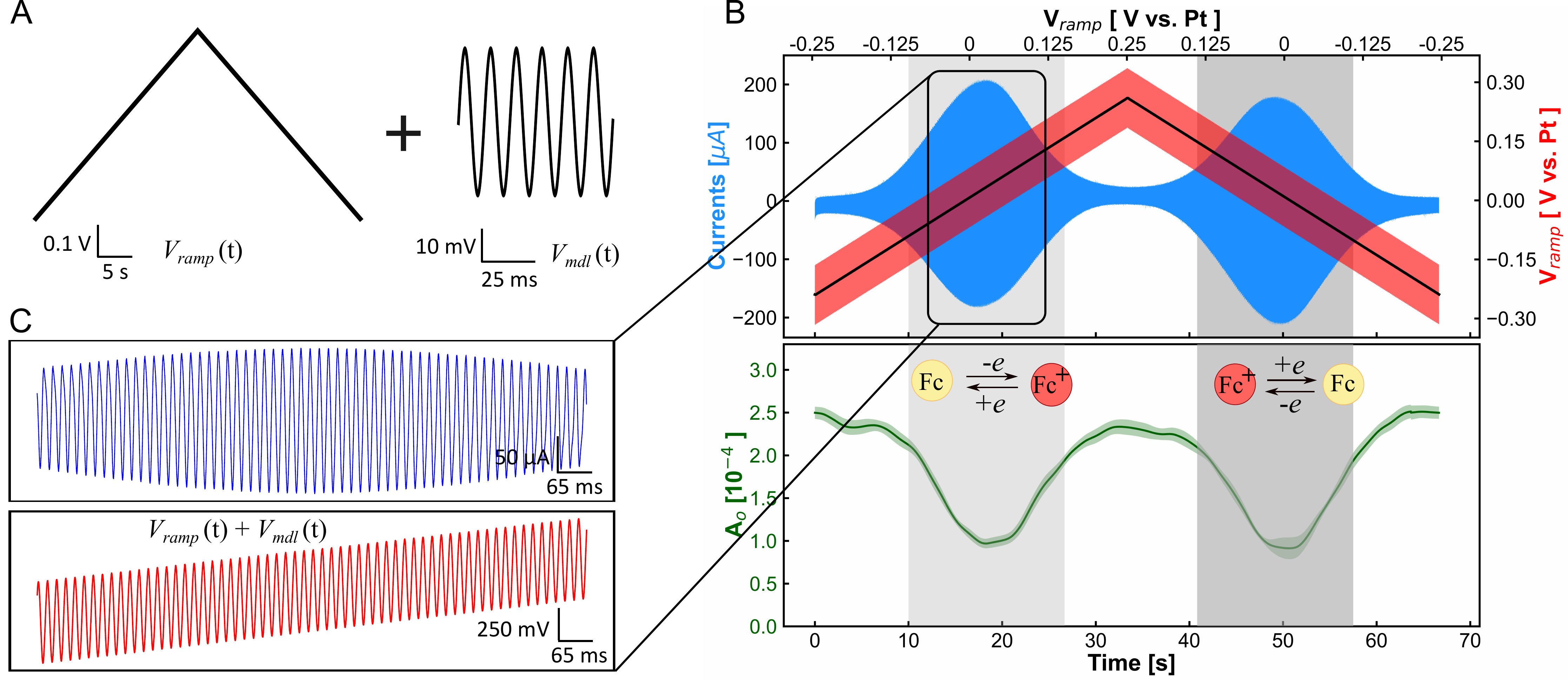}
\caption{\label{fig_Ave_ACV} The AC voltammetry measurements. (A) The applied AC potential is the superposition of an AC sinusoidal modulation potential and a linear triangle offset potential. (B) Top panel: The applied modulation potential and quasi-DC scan potential(red curve) and the response currents from the nanohole electrode(blue curve). Bottom panel: The averaged amplitude of the modulated optical signal over 8 cycles. Filled areas are the standard deviation of the averaged points. (C) Zoom-in on the applied potential and currents. The scan rate of the quasi-DC potential is $15$~mV/s from $-0.25$~V to $0.25$~V. The amplitude of the potential modulation is $50$~mV. Electrolyte solution is $10$~mM~\ch{Fc(MeOH)2} in $0.1$~M~\ch{KCl} solution. }
\end{figure*}

Since the combination of potential modulation and lock-in detection can provide much higher sensitivity, we have extended this approach to monitor the redox reaction of~\ch{Fc(MeOH)2} inside the nanoholes.
Instead of applying a linear scanning potential to the nanohole electrode (as in conventional cyclic voltammetry), we add a modulation to the scanning potential. In conventional amperometric measurements, this method is known as Alternating Current Voltammetry (ACV)~\cite{ElectrochemicalMethodsFundamentals, baranska_practical_2024, gavaghan_complete_2000, zhang_large-amplitude_2004}.
In ACV measurements, we call the linear potential scan a quasi-DC potential scan, the potential modulation is an AC potential, and their corresponding currents are labeled quasi-DC currents and AC currents, respectively.
As shown in Fig.~\ref{fig_Ave_ACV}A (not to scale), the applied potential is the superposition of a high-frequency small-amplitude sinusoidal AC potential with an amplitude of $50$~mV and frequency of $75$~Hz, and a linear triangle offset potential altering from $-0.25$~V to $+0.25$~V with a slow scanning rate of $15\text{ mV/s}$. 
We measure the time-dependent response of the light scattering intensity in a single nanohole, where the amplitude of the optical oscillations is detected accurately by the lock-in amplifier.
Here we note that the quasi-DC scan rate $15\text{~mV/s}$ is identical to only $0.006 \frac{k_BT}{e}/\tau_{RC}$, where $k_BT/e\simeq 25\text{~mV}$ is the typical (thermal) voltage scale of the EDL at room temperature and $\tau_{RC} = \lambda _{d} L/D \approx 5$~ms is the RC time of the system in terms of the Debye length $\lambda_d\simeq 1\text{~nm}$, the distance $L\simeq10\text{~mm}$ between the nanohole and the counter electrode, and the typical ion diffusion coefficient $D\simeq2\times10^{-9}\text{~m$^2$/s}$. In other words, the quasi-DC scan rate is much slower than the typical EDL charging rate, and hence we can indeed safely consider it as a steady-state potential.  In contrast, the RC time is shorter than (but of the same order as) the period $1/f=13.3\text{~ms}$ of the AC modulation frequency $f=75\text{~Hz}$, such that significant (although incomplete) charging and discharging of the EDL is to be expected during a period. 
%is larger than the equivalent frequency of the RC time of the system (200~Hz at $\tau_{RC} = 5$~ms).   

%The black curve in Figure S3(c) represents the light scattering intensities collected by PD during the ACV measurement.
%It is difficult to distinguish changes or any trend with the eye from the PD signal because the modulation potential $45$~mV is relatively small and the data are not averaged over many cycles.
%With the lock-in amplifier, we can get the corresponding amplitude of the PD signal.
%This observation demonstrates that we can use opto-iontronic microscopy to measure the information that is contained in EDL charging/discharging and electrochemical reactions via this optical signal.
%In the potential window (grey highlighted area in Fig.~\ref{fig_Ave_ACV}(b) ) of the electrochemical reactions, the amplitude of the optical signal decreases to the base value (around $0.7\times 10^{-4}$).

In Fig.~\ref{fig_Ave_ACV}B, we show the applied potential (black and red) and the resulting current (blue) in the upper panel, and the amplitude $A_0$ of the optical intensity in the lower panel, where we note that $A_0$ is averaged over 8 quasi-DC cycles of the ACV measurements. The inset in Fig.~\ref{fig_Ave_ACV}C  shows a magnification of the potential and the current. 
The quasi-DC potential scan starts at time $t=0$ from $-0.25$~V, which is such that only high-frequency capacitive EDL charging/discharging takes place inside the nanohole. The amplitude of the modulating optical signal stays constant at a relatively high level of about $A_0\simeq 2.5\times10^{-4}$. Compared to the EDL charging/discharging measurements shown in Fig~\ref{fig_DC_CV}A, the modulation potential amplitude in the ACV measurements is only ~50mV, which is 8 times lower, yet the noise in $A_0$ over the 8 cycles is small which demonstrates the reproducibility, stability, and sensitivity of the opto-iontronic microscopy (more details are shown in Fig. S4 $SI~Appendix$).
%In this EDL regime, the charging and discharging behavior can be modeled as an RC circuit.  The impedance of the RC circuit is $Z = R_{s} + \frac{1}{j \omega C_{EDL}}$. The characteristic charging time of the electrical double layer is given by $\tau_{d} = \lambda  _{d}^{2}/D \approx 1$~ns, $D$ is the diffusion constant of the ions \ch{K+}, \ch{Cl-}, \ch{Fc} and \ch{Fc+}, and $\lambda  _{d}$ is the Debye length, which is $1$~nm at $100$~mM \ch{KCl} solution. The RC time constant of the system is $\tau_{RC} =  R_{s} C_{EDL} = \lambda _{d} L/D \approx 5$~ms, where L is the distance between the nanohole electrode to the ground, which is~10~mm. 
%In Figure S2 in $SI~Appendix$, we confirm this crossover frequency $f_{co}$ by identifying the point at which the phase of the current response crosses $45\textdegree$ in the plot of phase and amplitude as a function of modulation frequency. We further estimate here $RC = 1/(2\pi f_{co}$) with $f_{co} =200$~Hz, with the $C_{EDL}=1.25~\mu\text{F}$, we thus would get $R_{S}= 0.63~\text{k}\Omega$.
%In contrast, the EDL on the surface of the nanohole has only a short time to respond to the high-frequency ($f >$ 100~Hz) modulation potential, resulting in a reduced optical signal. 
At about $t=8\text{~s}$, the potential window of the ~\ch{Fc(MeOH)2} redox reaction is entered, which gives rise to a large increase by an order of magnitude of the current amplitude (blue). The current in this window peaks at about $t=18\text{ s}$ at a voltage of around $+50$~mV, which, interestingly, coincides with a significant change of the optical signal, with a drop by a factor 2.5 to $A_0\simeq 1.0\times10^{-4}$.
% Similarly, the electrochemical reaction system can be modeled as an RC circuit, as illustrated in Fig~\ref{fig_DC_CV}(e).  The impedance of the RC circuit is $Z_{ec} = R_{s} + \frac{R_{ct}}{1 + j \omega R_{ct} C_{EDL}}$. Then we can get the RC time of the reaction interface $\tau_{ec} = R_{ct} C_{EDL}$. The crossover frequency is $f_{co} = \frac{1}{1/( 2\pi R_{ct} C_{EDL}}$, with the $C_{EDL}$ is already known as $1.25~\mu F$, and from the phase response of the cuurent during the electrochemical reaction with the function of potential frquency, we can get the $f_{co}$ at the crossover frequency $f_{co}$ by identifying the point at which the phase of the current response crosses $45\textdegree$ in the, which is  $f_{co}$ = xxx Hz, see more details in the FigureS4 in SI. Hence we can get the $R_{ct}=xxx~\text{k}\Omega$.
The potential is increased further to reach the maximum of $0.25$~V at $t=33\text{ ~s}$, when it is out of the redox window of ~\ch{Fc(MeOH)2}.  
As the quasi-DC potential sweeps outside of the redox window at about $150$~mV, the amplitude of the current goes down while the amplitude of the modulated optical signal recovers again, returning to about $A_0\simeq 2.5\times10^{-4}$.
% The phase of the modulated optical signal goes back to 170\textdegree~as well.
At $t=33\text{~s}$ the quasi-DC potential reverses and scans from $0.25$~V back to $-0.25$~V. Again, when the quasi-DC potential is in the window of the redox reactions, the current increases by an order of magnitude and the amplitude of the modulated optical signal drops to $A_0=1.0\times10^{-4}$. Finally, the current and  $A_0$ return to their original values when the quasi-DC potential leaves the potential window of the redox reactions. We also recorded the phase difference between the applied voltage and the optical signal, however its variation is much less significant than that of the amplitude and will only be discussed briefly in Fig. S5 ($SI~Appendix$).  
%could be affected by the difference of the diffusion constants between the \ch{Fc(MeOH)2} and \ch{K+}, \ch{Cl-} ions, and the effects are relatively small compared to the amplitude changes, hence we focus on the amplitude changes.  As a reference, in Figure S6 in $SI~Appendix$, we show the amplitude and phase information of the modulated light scattering intensities stay at a constant level during the ACV measurement without~\ch{Fc(MeOH)2} in the supporting electrolyte of $100$~mM~\ch{KCl}. 

%The information we can get from Fig.~\ref{fig_Ave_ACV} is that when the potential is modulated out of the potential window of redox reaction of~\ch{Fc(MeOH)2}, there is only EDL charging and discharging inside the nanohole. The amplitude of the modulated optical signal stays constant since the amplitude of the potential modulation is constant at $50$~mV. While the potential is modulated in the potential window of redox reactions, there are both EDL charging/discharging and redox reactions inside the nanohole.

\begin{figure*}[htbp]
\centering\includegraphics[width=\textwidth]{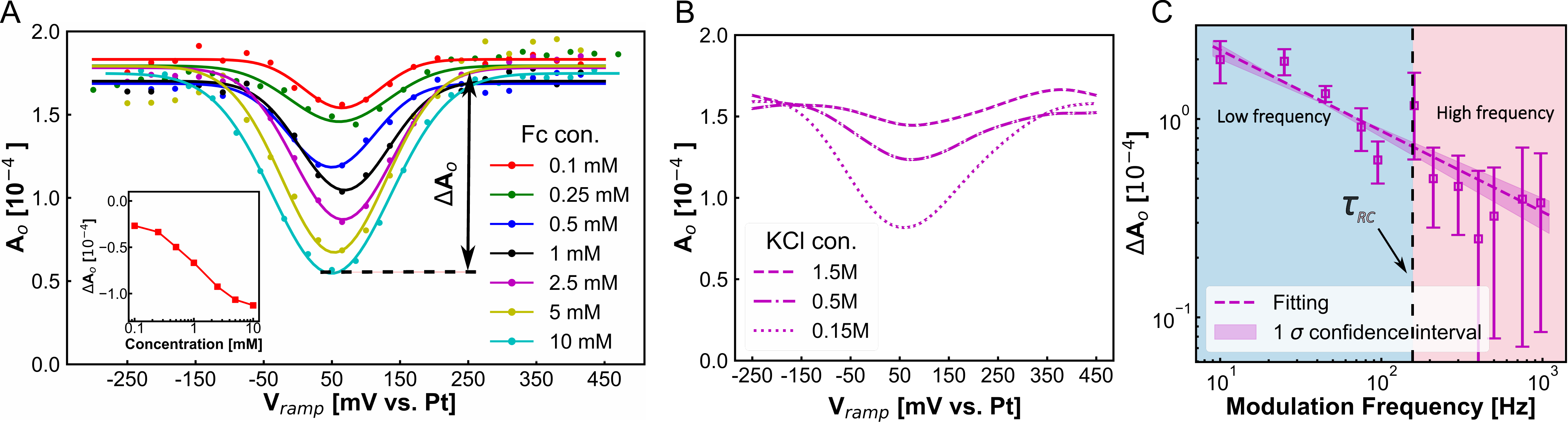}
\caption{\label{DifferentFc} (A) The amplitude of the modulated optical signal in the function of potential in different concentrations of~\ch{Fc(MeOH)2} from 0.1~mM to 10~mM. The supporting electrolyte keeps at the same concentration at $0.1$~M~\ch{KCl}. The amplitude of the modulation potential is $V_{0}= 100$~mV, the frequency of the modulation potential is $f = 975$~Hz, the step of the quasi-DC potential scan is $V_{step}= 35$~mV, with the period of $t_{step}= 100$~s for each step. (B) The amplitude of the modulated optical signal as a function of AC potential in different concentrations of supporting electrolyte~\ch{KCl} from 0.15~M to 1.5~M. The concentration of~\ch{Fc(MeOH)2} remains the same at $2$~mM. The amplitude of the modulation potential is $V_{0}=$ 100~mV, the frequency of the modulation potential is $f =$ 975~Hz, the quasi-DC potential linearly scans from $-250$~mV to $450$~mV and then scans back at a scan rate of $R_{scan}= 50$~mV/s. (C) The amplitude difference $\Delta A_{0}$ of the modulated optical signal as a function of the modulation frequency from 10~Hz to 975~Hz, the amplitude of the modulation potential is $V_{0}= 100$~mV, the concentration of ~\ch{Fc(MeOH)2} is $2$~mM. }
\end{figure*}

We conclude from Fig.\ref{fig_Ave_ACV} that the amplitude $A_0$ of the modulation of the optical intensity as measured from a single nanohole during ACV is strongly correlated to the electric current. This current only contains a small (constant) capacitive component outside the redox window and an additional (large) Faradaic component inside the redox window. The large decrease of $A_0$ in the redox window strongly suggests that it is caused by a relatively large concentration change of (one of) the redox species inside the nanohole, be it \ch{Fc(MeOH)2+} or \ch{Fc(MeOH)2} or both. 
%The amplitude of the modulated optical signal goes down since there is an ion flux of~\ch{Fc(MeOH)2+} at the inner surface of the nanohole, which affects the local ion concentration inside the nanohole.
In order to investigate this hypothesis, we measured $A_0$ for seven bulk concentrations of ~\ch{Fc(MeOH)2} between $0.1$~mM and $10$~mM, for the same background electrolyte concentration and the same scanning potential as in Fig.\ref{fig_Ave_ACV}. 
In Fig.~\ref{DifferentFc}A we show the resulting amplitude $A_0$ of the optical intensity as a function of the quasi-DC potential, revealing a drop in the redox window for all seven ~\ch{Fc(MeOH)2} concentrations, the drop being larger at higher concentrations of redox species. This is qualitatively in line with our hypothesis. The inset quantifies the change $\Delta A_0$, defined as the difference between the minimum of $A_0$ in the redox window and its value far outside, and reveals a variation from $\Delta A_0=-0.2\times 10^{-4}$ at the lowest concentration to $\Delta A_0=-1.2\times 10^{-4}$  for the highest concentration of ~\ch{Fc(MeOH)2}. 

%The amplitude of the modulated optical signal starts at a level of $1.75\times10^{-4}$ and goes down with the quasi-DC potential in the redox potential window.
%We see a reduction of the amplitude in the redox window, which becomes more pronounced with higher~\ch{Fc(MeOH)2} concentration.
%a higher concentration of~\ch{Fc(MeOH)2} inside the nanohole, more~\ch{Fc(MeOH)2} molecules will contribute to the redox reactions inside the nanohole in the redox potential window.
%As a consequence, one expects large ion concentration variations and large currents inside the nanohole. 
%This picture is considered with the observation that the change of the concentration of~\ch{Fc(MeOH)2} only affects the amplitude of the optical signal at the redox window but has almost no effect on the amplitude of the optical signal outside of the reactivity potential window.
%Note that the amplitude change, $\Delta A_{O}$, is linearly proportional to the applied modulation amplitude, $V_{\mathrm{mdl}}$.

%In order to check which ion's concentration contributes most to the amplitude of the modulated optical signal and to the changes in the amplitude of the modulated optical signal during the redox reactions, 
We further investigated the dependence of $A_0$ on the concentration of the supporting electrolyte~\ch{KCl}, keeping the \ch{Fc(MeOH)2} concentration fixed at $2$~mM. For the three \ch{KCl} concentrations $0.15$, $0.5$, and $1.5$~M, the resulting relation between $A_0$ and the quasi-DC potential is 
%In the measurements, we keep the~\ch{Fc(MeOH)2} at a constant concentration at $2$~mM and change the supporting electrolyte concentration from $0.15$~M to $1.5$~M.
%For each concentration of the~\ch{KCl}, we do the AC voltammetry for the nanoholes, and the amplitude of the optical signal in the function of quasi-DC potential is 
shown in Fig.~\ref{DifferentFc}B. In the potential window of capacitive EDL charging-discharging, we see that $A_0\simeq1.6\times10^{-4}$ for all three \ch{KCl} concentrations, however in the potential window of additional redox reactions we observe a larger decrease of $A_0$ for lower salt concentrations. We interpret this qualitatively in terms of the relatively larger contribution of \ch{Fc(MeOH)2+} to the EDL at low \ch{KCl} concentrations. 
%At the redox potential of~\ch{Fc(MeOH)2}, the amplitude of the optical signal goes down more with lower~\ch{KCl} concentration, but the 
%These interesting results indicate that with lower~\ch{KCl} concentration, the screen length is longer and the~\ch{Fc(MeOH)2} contributes relatively more to the formation of the EDL, hence induces more changes of the optical signal.
The electrochemical redox currents for different concentrations of~\ch{KCl} with 2~mM ~\ch{Fc(MeOH)2} are at the same level, see Fig. S7 in $SI~Appendix$.  

%By calculating the difference of the optical amplitude in and out of the redox window, we calculated the amplitude changes $\Delta$A$_{O}$, 
% Fig.~\ref{DifferentFc}(c) shows the optical amplitude changes $\Delta$A$_{O}$  as a function of the modulation amplitude. The amplitude changes of the optical signal linearly increase with the increase of the modulation amplitude from $20$~mV to $150$~mV.
% This indicates that with higher modulation amplitude, the number of redox~\ch{Fc} molecules will take place inside the nanohole, hence the optical changes caused by the conversion~\ch{Fc}/\ch{Fc+} increase more.
Fig.~\ref{DifferentFc}C shows the optical amplitude changes $\Delta$A$_{0}$ as a function of the modulation frequency.  The amplitude changes of the optical signal decrease with the increase of the modulation frequency from $10$~Hz to $975$~Hz.
Below the crossover frequency 
$f_{CO} = 1/\tau_{RC}$, the amplitude changes $\Delta$A$_{0}$ shows a gradual decrease with frequency, consistent with the low-frequency regime where the system can follow the modulation efficiently. When the modulation frequency $f $ is above $f_{CO}$ the $\Delta$A$_{O}$ exhibits larger standard deviations, indicating that the optical response is strongly attenuated due to the limited RC charging dynamics, as reflected by the fit and its confidence interval.

\section{Modeling the redox reactions inside a nanohole}
\begin{figure*}[ht]
\centering\includegraphics[width=\textwidth]{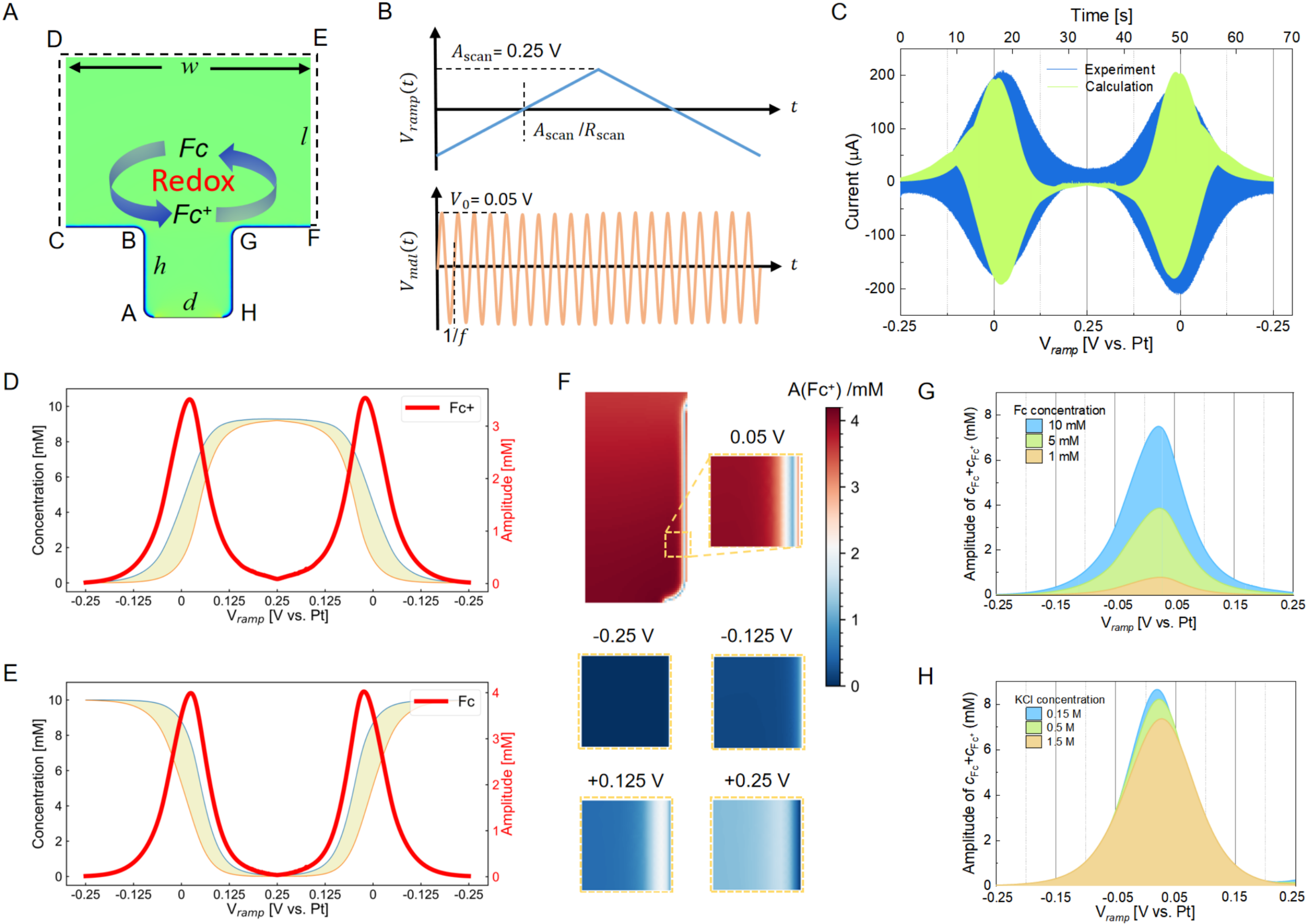}
\caption{\label{Model} Results obtained from numerical solutions of the Poisson-Nernst-Planck and Butler-Volmer (PNP-BV) equations for diffusion and migration coupled to the redox reaction. (A) Schematics (not to scale) of the geometry of a single cylindrical nanohole electrode in contact with the grounded bulk electrolyte, with $d = 100$~nm, $h = 100$~nm, $w = 10~d$, and $l = 1000~h$. (B) Time-dependence of the slow triangular quasi-DC scanning potential (peak values $\pm250$~mV and scan rate $15\text{~mV/s}$) and its superimposed fast small-amplitude sinusoidal modulation (amplitude $50\text{~mV}$ and frequency $ 75\text{~Hz}$). (C) Comparison between the PNP-BV calculation (green) and experiment (blue) for time-dependent current at bottom point ``H'', 
shown for bulk concentrations $C_{\mathrm{Fc}}$ = $10$ mM and $C_{\mathrm{KCl}}$ = $0.1$ M. The calculated current density $j$ was multiplied by the total surface area of the electrode $A_{\mathrm{tot}}$. Time-evolution (yellow) of the nanohole-averaged concentration of (D)~\ch{Fc+} and (E)~\ch{Fc} inside the nanohole for $C_{\mathrm{Fc}}$ = $10$ mM and $C_{\mathrm{KCl}}=0.1$ M, with their maxima (blue) and minima (orange). Their difference is the concentration modulation amplitude (red). (F) 2D heat map of the local ~\ch{Fc+} concentration modulation amplitude (mM) for $C_{\mathrm{Fc}}$ = $10$ mM and $C_{\mathrm{KCl}}$ = $0.1$ M. Sum of the concentration modulation amplitude of \ch{Fc} and \ch{Fc+} as a function of the scanning potential, for various bulk concentrations of (G) \ch{Fc} with $C_{\mathrm{KCl}}$ = $0.1$ M, and (H) \ch{KCl} with $C_{\mathrm{Fc}} =10$ mM.}
\end{figure*}

We theoretically model the reaction-transport processes in a single nanohole to explore how the modulation potential affects the distribution of species inside the nanohole, with the goal of identifying the key species that influence the optical contrast measurements. We employ the standard Poisson-Nernst-Planck (PNP) equations for ionic diffusion and migration and couple them to the Butler-Volmer (BV) equation for the redox reaction, as introduced in the Methods section. 
We expect rectification effects to be relatively small in the present system because the Debye length of 10 nm is an order of magnitude smaller than the typical linear length scales of the nanoholes. In principle this rectification effect is included in the PNP equations, although a full analysis of rectification would require a coupling to the Stokes equation for electro-osmotic flow contribution.
In the model geometry shown in Fig.~\ref{Model}A, a cylindrical nanohole with diameter $d=100$~nm and height $h=100$~nm is connected to a large cylindrical reservoir of width $w=10d$ and length $l=1000h$ containing the redox species~\ch{Fc(MeOH)2} and the aqueous support electrolyte~\ch{KCl} at given bulk concentrations $C_{\mathrm{Fc}}$ and $C_{\mathrm{KCl}}$, respectively. Under an applied electrode potential $V(t)$, a reversible redox reaction $\mathrm{Fc} \rightleftharpoons \mathrm{Fc^+}+e^{-}$ occurs at the electrode-electrolyte interface, where \ch{Fc} and \ch{Fc+} are short for~\ch{Fc(MeOH)2} and~\ch{Fc(MeOH)2+}, respectively.  
%\begin{equation}
%\mathrm{Fc} \rightleftharpoons \mathrm{Fc^+}+e^{-}
%\end{equation}
As shown in Fig.~\ref{Model}B, the electrode potential $V(t)=V_{ramp}(t)+V_{mdl}(t)$ is composed of (i) a slow triangular quasi-DC scanning potential $V_{ramp}(t)$ with peak values $A_{scan}=\pm250$~mV and  scan rate $R_{scan}=15\text{~mV/s}$  such that the period is 66.7~s and (ii) a sinusoidal potential $V_{mdl}(t)=V_0\sin(ft)$ with amplitude  $V_0=50\text{~mV}$ and frequency $f =75$~Hz.
%which is expressed as
%\begin{equation}
%\phi(t)=\phi_{DC}(t)+\phi_{AC}(t)
%\end{equation}
%The peak values of the quasi-DC scanning potential are $A_{scan}=\pm250$~mV, and different scan rates are chosen as $V_{scan}$ = 25 or 50~mV/s. The modulation amplitude of the sinusoidal potential is $A_{mdl}=50$~mV and the modulation frequency is $f_{mdl}=8$~Hz.

%RvR: The peak values of $V_{ramp}(t)$ are $\pm250$~mV and with its slow scan rate of 25~mV/s the period of the quasi-static DC potential equals 20~s. The sinusoidal modulation potential $V_{mdl}(t)$ has an amplitude of $50$~mV and a frequency $f=8$~Hz.  

The green curves in Fig.~\ref{Model}(C) show the resulting PNP-BV prediction for the time-dependent reduction current density at the bottom corner point ``H'' of the nanohole (multiplied by the total electrode area $A_{\mathrm{tot}}=2.5\text{~mm}^2$) for the case of bulk concentrations $C_{\mathrm{Fc}}=10$~mM and $C_{\mathrm{KCl}}=0.1$~M. %$A_{scan}$ = $\pm250$~mV, $V_{scan}$ = $50$~mV/s, $A_{mdl}=50$~mV and $f_{mdl}$ = 8~Hz. 
Here we checked that the current density is rather uniform in the nanohole and that the whole macroscopic area $A_{\mathrm{tot}}$ contributes to the current. We observed for $t\in[0,8]\cup[24,40]\cup[56,66.7]$~s, that the amplitude of the current oscillations induced by the sinusoidal modulation voltage is small because the applied potential is outside the Fc redox potential window (between $-125$~mV and $125$~mV); in this regime only the charging/discharging of the EDL occurs. With the increase of the ramp voltage, the redox window is crossed between $t\in[8,24]\cup[40,56]\text{ s}$ during which reactions gradually occur and disappears again, exhibiting peaks at $t\simeq \text{18 and 49 s}$ where the current amplitude is an order of magnitude larger than outside the redox window.
%resulting in the two symmetric oscillation peaks at $\phi$(t = 12 s) = 0.05 V, substantial current oscillations are observed during $t\in[5,12]$~s, with amplitudes that are an order of magnitude larger than outside the redox potential window. 
%This is because the amplitude of the sinusoidal modulation voltage matches the redox potential window (between $-125$~mV and $125$~mV), causing substantial redox current oscillations. 
We conclude, and further confirm from the blue curves in Fig.~\ref{Model}C, that the AC-CV curves obtained from numerical calculations are very similar to the experimental results (Fig.~\ref{fig_Ave_ACV}B), which provides confidence in the model and its parameters. 
One of the main advantages of having this consistent numerical model is the direct access that it provides to the time-evolution of the concentration profile of each dissolved species inside the nanohole, quantities that are not directly accessible experimentally. 

Fig.~\ref{Model}(D) and (E) depict the model predictions of the time-dependent average concentrations of~\ch{Fc+} (D) and~\ch{Fc} (E) inside the nanohole, for the same system parameters as in C, as a function of the quasi-static scanning voltage $V_{ramp}(t)$. The two concentrations reveal high-frequency (75 Hz) oscillations due to the modulation potential, which are represented by the yellow hatch as the oscillations cannot be resolved on the scale of (D) and (E). The blue and orange curve represent the maximum and minimum of the concentration modulations, respectively, and reveal that 
%system parameters are $C_{\mathrm{Fc}}$ = $10$ mM, $C_{\mathrm{KCl}}$ = $0.1$~M and $R_{scan}$ = $15$~mV/s. 
the redox reaction continuously produces  \ch{Fc+} when the quasi-DC scanning voltage varies from $-250$~mV to $+250$~mV during time $t\in[0,33.3]$~s, while the reactant~\ch{Fc} is gradually consumed. Upon scanning back from $+250$~mV down to $-250$~mV, the reverse is predicted to occur, with a production of \ch{Fc} at the expense of the consumption of \ch{Fc+}. 
To further identify and analyze the key species that affect the measured optical signal, we extract the response amplitude of the concentration of each species to the modulation potential from the calculated concentration distribution. 
The red lines in Fig.~\ref{Model}(D) and (E) show the time-dependent amplitude of the high-frequency concentration modulation of (D) \ch{Fc} and (E) \ch{Fc+},
%at bulk concentrations  $C_{\mathrm{Fc}}$ = 10 mM, $C_{\mathrm{KCl}}$ = $0.1$~M and scan rate $R_{scan}$ = $15$~mV/s, 
revealing a mutual exchange during the voltage sweep through the redox window with a concentration decrease/increase of \ch{Fc}/\ch{Fc+} due to the oxidation-dominated regime during $t\in[0,33.3]\text{ s}$ and vice versa during the reduction-dominated regime during $t\in[33.3,66.7]\text{ s}$. 
Clearly, distinct peaks appear in the time-dependent concentration response amplitudes of~\ch{Fc} and~\ch{Fc+} near the redox potential window. 

In Fig.\ref{Model}(F) we present, for several ramp voltages, 2D heat maps of the local ~\ch{Fc+} concentration modulation amplitude $A$(\ch{Fc+}). The overall homogeneity of these heat maps
shows a spatial extent of the concentration modulations far beyond the range of the EDL.  This homogeneity on the length scale $h=d=100\text{~nm}$ of the nanohole is consistent with the typical diffusion time $h^2/D\simeq 10^{-5}\text{~s}$ of ions and molecules in the hole. This timescale is orders of magnitude shorter than the period of the 75 Hz modulation potential, such that essentially complete mixing of the produced and consumed redox species over the nanohole is achieved during each modulation cycle. 
%indicate that the modulation potential induces the rapid cycling of redox species in the nanohole that amplifies the current signal in the redox potential window. 
Significant heterogeneity of the concentration amplitudes is only observed in the vicinity of the EDL close to electrode-electrolyte interface, especially 
%at We also see that $A$(\ch{Fc+}) displays differently between the EDL region and the bulk in-hole space 
at $V_{ramp}=+0.05\text{~V}$ where the relatively small modulation amplitude $A$(\ch{Fc+}) in the EDL region is attributed to the positive quasi-DC scanning potential that expels produced \ch{Fc+} and alleviates the cycling of \ch{Fc+} in the EDL. By contrast, we observe that the ion concentrations of the supporting electrolyte (\ch{K+} and~\ch{Cl-}) depend mainly on the charging/discharging of the EDL, with maximum amplitude occurring at higher positive or negative potentials and no significant enhancement of the response within the redox potential window (Fig. S10-12 in $SI~Appendix$). Therefore, we can conclude that the change in EDL-modulation contrast inside the reactive potential window must be caused by the significant oscillations of the concentrations of~\ch{Fc} and~\ch{Fc+}. This observation further highlights the applicability of opto-iontronic microscopy for monitoring electrochemical activity.

To further study the contribution of each species to the amplitude of the optical modulation, we investigate the effects of different bulk concentrations of \ch{Fc} and \ch{KCl} on the potential-dependence of the response amplitudes of \ch{Fc} and \ch{Fc+}. 
In Fig.~\ref{Model}(G), where we set $C_{\mathrm{KCl}}=0.1\text{~M}$ as before, we plot the sum of the predicted concentration modulation amplitudes $A$(\ch{Fc})+$A$(\ch{Fc+}) for the three bulk \ch{Fc} concentrations $C_{\mathrm{Fc}}=1\text{~mM}$ (orange), 5~mM (green), and 10~mM (blue). We see, as expected, that all three curves peak in the middle of the redox window, with a peak value that equals essentially $70-80\%$ the bulk concentration $C_{\mathrm{Fc}}$. Thus, the dependence of the cycling of the redox species on $C_{\mathrm{Fc}}$ involves, at least in the concentration regime studied here, a simple linear scaling relation.    
%increases from $1$~mM to $10$~mM, the response amplitudes of total concentrations of~\ch{Fc} and~\ch{Fc+} increased in the window of the redox potential, as expected. 
In Fig.~\ref{Model}H, where we set $C_{\mathrm{Fc}} = 10$~mM as before, we plot voltage dependence of $A$(\ch{Fc})+$A$(\ch{Fc+}) for the three \ch{KCl} bulk concentrations $C_{\mathrm{KCl}}=0.15$~M (blue), 0.5~M (green), and 1.5~M (blue). The dependence on $C_{\mathrm{KCl}}$ is relatively weak yet systematic, where we attribute the somewhat lower redox efficiency at higher salt concentration to the enhanced electrode screening at higher $C_{\mathrm{KCl}}$.  
%increases from $0.15$~M to $1.5$~M, the response amplitudes of total~\ch{Fc} and~\ch{Fc+} concentrations at the redox potential window gradually decreases, which is also expected on the basis of enhanced screening (and hence less redox efficiency) at higher salt concentrations. 
In our experiments, as the bulk \ch{Fc} concentration increased from $0.1$~mM to $10$~mM (Fig.~\ref{DifferentFc}A), the amplitude of the modulation of the optical signal decreased within the redox potential window (between $-125$ and $125$~mV). Qualitatively this stronger optical response at higher $C_{\mathrm{Fc}}$ is consistent with our theoretical predictions of a larger amplitude of the concentration modulations of the redox species at higher $C_{\mathrm{Fc}}$. Quantitatively, however, there is a mismatch between the linearity of the theoretically predicted $C_{\mathrm{Fc}}$-dependence of the concentration amplitudes and the logarithmic $C_{\mathrm{Fc}}$-dependence of the measured optical intensity (see inset of Fig.\ref{DifferentFc}(A)). This requires further investigation. Also the dependence of the measured optical signal on KCl concentration, shown in Fig.~\ref{DifferentFc}(B) for three salt concentrations, is in qualitative agreement with our predictions, however quantitatively the mapping between the measured optical signal and the calculated concentrations of redox species remains elusive.  With the~\ch{KCl} concentrations  $0.15$~M to $1.5$~M, in (Fig.~\ref{DifferentFc}B), the amplitude of optical modulation at the EDL charging potential window ($\pm$250 mV) remained around $1.6\times 10^{-4}$, further proofing that KCl concentration is not a key factor affecting the amplitude of the optical modulation. However, lower~\ch{KCl} concentrations led to a more apparent decrease in the amplitude of the optical modulation at the redox potential window, suggesting that lower supporting electrolyte concentrations amplify the influence of~\ch{Fc} on signal changes of the optical modulation.

\section{Conclusion}
We have developed an opto-iontronic microscopy technique utilizing potential modulation and lock-in detection that significantly enhances the sensitivity of measurements of electric double layer (EDL) charging/discharging and electrochemical reactions in an aqueous electrolyte. 
We have fabricated an electrolyte-immersed electrode containing a large number of cylindrical nanoholes (diameter $75$ nm and depth $100$ nm) that each contain an extremely small electrolyte volume in the attoliter regime. Such a small volume, and hence a relatively large area of the electrode-electrolyte interface, allows for the optical detection of composition changes in the nanohole driven by an applied electrode potential. We showed here that these changes are not only optically observable for purely capacitive EDL charging/discharging processes in a chemically inert support electrolyte, but also for Faradaic processes in oxidation-reduction reactions.
In particular, we have demonstrated the versatility and sensitivity of opto-iontronic microscopy by its ability to monitor the electrochemical redox reaction of 1,1-Ferrocenedimethanol (abbreviated as \ch{Fc}) inside a single nanohole during Alternating Current Voltammetry (ACV) experiments with an applied electrode potential $V(t)$ that consists of the sum of a slow scanning potential and a fast (high-frequency) low-voltage modulation.
In order to prove that the observed changes in the optical contrast during the redox reaction are indeed due to the concentration changes of the reduced (\ch{Fc}) and oxidized (\ch{Fc+}) form of the redox couple, we conducted numerical calculations based on the Poisson\nobreakdash--Nernst\nobreakdash--Planck (PNP) and Butler\nobreakdash--Volmer (BV) equations to model ion transport and the redox reaction within the nanohole during the electrochemical reaction. These calculations, which accounts for reaction-diffusion-migration processes driven by a time-dependent applied electrode potential $V(t)$, not only predict the measured electric current $i(t)$ in good agreement with the experiments, but they also yield the spatially resolved time-dependent ion concentration profiles $C_i({\bf r},t)$ of all species inside the nanohole during the ACV-induced redox reactions of \ch{Fc}. Our calculations show that the concentrations of the two redox species \ch{Fc} and \ch{Fc+}, averaged over the nanohole volume, vary enormously during the quasi-static crossing of the redox window, cycling with the modulation frequency between a minumum and a maximum with concentration amplitudes exceeding 50\% of the bulk \ch{Fc} concentration. By contrast,  the background \ch{KCl} concentrations vary typically by only 10\% throughout the redox window. This is clear evidence that the signal we detect through opto-iontronic microscopy is primarily sensitive to the concentration of the redox species. Future research should focus on the improvement of the detection sensitivity by reducing the noise of the system and also focus on the quantitative characterization of the relation between the optical signal and the electrochemical processes with spectroscopy.

Although we focus here on a model system, the approach is general because it tracks interfacial electrochemical changes through an optical readout under electrochemical control. In principle, it can be extended to other reactions that modulate local ion concentrations and pH near the interface of electrolyte-electrodes, such as hydrogen evolution reaction (HER), oxygen evolution reaction (OER) as well as \ch{CO2} reduction reactions (CO2RR) and redox reactions in flow batteries. 
Furthermore, the proposed approach can also be applied to characterize the chemical states of the electrode materials. This can be further applied to monitor electrode/catalyst states, including corrosion/passivation (Pt cathode corrosion during HER), catalyst transformation (Cu/\ch{Cu2O} during \ch{CO2} reduction), ions intercalation in electrode (Li ion intercalation in electrodes), or metal deposition/stripping (Pt, Pd, Ni catalyst deposition on electrode). Importantly, these extensions primarily require adapting the electrochemical protocol and the calibration rather than changing the core measurement platform, enabling broad applicability to nanoscale electrochemical processes.

\begin{acknowledgments}
This research was supported by the Netherlands Organization for Scientific Research (NWO grant 680.91.16.03). H.T. thanks the support of National Natural Science Foundation of China (22508113). All authors thank Paul Jurrius, Dave van den Heuvel, and Dante Killian for technical support and fruitful discussions. Z.Z. thanks the Electron Microscopy Center at Utrecht University for the training and use of SEM-FIB system, and Z.Z. also thanks the AMOLF NanoLab Amsterdam for the training and use of EBPVD system. 
\end{acknowledgments}
% \vspace{0.5 cm}
\section*{Author contributions}
Author contributions: Z.Z. conceptualized the work and carried out the experiments; H.T. developed the theory; Z.Z. and H.T. analyzed data; C.L.\ contributed to developing the theory; R.v.R. and S.F. supervised the research.  All authors discussed the results and contributed to the manuscript.

\section*{Methods}
\subsection*{Setup}

Our microscope instrument works based on dark-field scattering from the glass interface under total-internal-reflection (TIR) illumination~ \cite{naminkElectricDoubleLayerModulationMicroscopy2020, meng_sensing_2021, mengMicromirrorTotalInternal2021}.
A laser beam is generated by an external-cavity diode laser (Ignis, $640$~nm, Laser quantum) with a laser controller (DPSS laser controller, Laser quantum) with a maximum output power of $300$~mW.
To improve the illumination quality, we use a telescope (T) with two plano-convex lenses arranged in a 4\textit{f} geometry to adjust the beam diameter to $2$~mm. 
The beam is then focused by a plano-convex lens and reflected by a small prism mirror (MM) ($5\times5$~ mm), which is off-axis of the optical path into the back focal plane (BFP) of a high-numerical-aperture (high-NA) microscope objective (Nikon, CFI Apochromat TIRF $60\times$, $1.49$ NA). 
Two mirrors M1 and M2 are used to precisely align the beam to the center of the small prism mirror and also ensure the reproducibility of the alignment.
Before the beam is directed toward the back aperture of the microscope objective, we use an adjustable mount to change the angle of the small prism mirror, which enables precise control of the incidence angle.
Since the laser beam is shifted from the optical axis of the high-NA ($1.49$) objective, the refracted light hits the interface of glass slides and water with an incident angle large enough to be reflected by TIR. 
The total-internal-reflected beam passes through the high-NA objective and is guided to the QPD. 
The illumination spot, where the TIR happens, has an elliptical shape of minor axis $100$~$\mu$m and major axis $220$~$\mu$m.  In this area, the evanescent wave on the interface between the glass side and electrolyte scatters from the surface inhomogeneities and all objects that are placed at the interface.

Part of the scattered light is collected by the high-NA objective and passed to a mirror (M3). 
The reflected light by M3 is focused onto a scientific complementary metal-oxide semiconductor (sCMOS) camera (ORCA-Flash 4.0 V3, Hamamatsu) with a lens (L3) and through a pinhole on a silicon photodiode (PD, $\Phi1.2$~mm Si-PIN, $320 - 1060$~nm ).
In Fig.~\ref{fig_Setup_SEM_CMOS}B, we show the scattering light image of a nanohole array in the whole illumination area. 
For potentiodynamic measurements, the required modulation frequency of the electrode potential is comparable to or higher than the reachable frame rate (fps) of the sCMOS camera, which is typically around $200$~fps for a customized field of view. 
The camera can achieve up to~$1000$~fps, but it is at the expense of a narrower field of view (FOV). 
To get high sampling rates of the scattered light from the objects, we employed a second image arm in the image module by adding a beamsplitter (BS) to reflect the light to the mirror (M4) partially, then the reflected light is filtered by a pinhole, which is used to select a single scattering light spot from the camera image. The selected scattered light spot is projected by a lens (L4) with 4\textit{f} geometry configuration on a Photo Diode (PD, PWPR-2K-SI, FEMTO). 
The PD enables us to have a high-frequency modulation measurement and a better signal-to-noise ratio (SNR), thanks to the ultra-low noise sampling of the PD.
The Point Spread Function (PSF) of the microscope is shown in $SI Appendix~ \text{Fig.~S14}$, obtained from scanning a selected nanohole through the pinhole.

We use a high-precision waveform generator (33120A, 15MHz, HP) or a Data acquisition card (DAQ)(NI USB-6212, National Instrument) to generate a modulation signal and a reference signal of the modulated signal.
The modulated signal is then sent to a potentiostat (E162 picostat, or EA362 Dual picostat,eDAQ), which accurately controls the cell potential during the experiments.
The reference signal is sent to the reference-in channel of the lock-in amplifier.  
The corresponding potential and current signal from the electrochemical cell are recorded and amplified by a data recorder (e-corder 410, eDAQ).
During the potential modulation, variation in the scattering light intensity from the investigated objects is collected by the PD, which converts light to current and amplifies the current to voltage signal (with a bandwidth of 2K and trans-impedance gain of $10^{9}~\Omega$, the rising time characterization of the PD is provided in Section Experimental details of the $SI~Appendix$). 
The amplified light intensity is sent to a lock-in amplifier (SR830 DSP lock-in Amplifier, Stanford Research System).
Based on the input signal of light intensity and the reference-in signal, the lock-in amplifier calculates the amplitude of the light intensity modulation caused by the potential modulation. The lock-in output signal is recorded by the computer through a DAQ and also connected to an oscilloscope for data monitoring.
More details about the lock-in amplifier parameter settings are provided in Section Experimental details of the $SI~Appendix$.
A four-channel oscilloscope (InfiniiVision DSOX2024A, Keysight) is used to monitor four signals in real time while the measurement is running. Two of the channels display the potential signal and current signal from the electrochemical cell.
One of the channels is used to monitor the scattering light intensity of a selected object from the PD.
One of the channels is connected to the output of the lock-in amplifier to monitor the amplitude of the light intensity variations. 

\subsection*{Data Acquisition}

We use a special-purpose \textit{Python} program to interface the camera, which allows us to grab and store image frames from the camera.
The program can communicate with the waveform generator, lock-in amplifier, potentiostat, and DAQ card. 
Before the measurement starts, we set all the measurement parameters (\textit{e.g.}, potential, scan rate of potential, frequency, amplitude of modulation, measurement time) through the program.
Then the measurement starts to run and all the electronics are synchronized by the DAQ card, which is also used to collect all the signals and store them in the PC by the \textit{Python} program. 

\subsection*{Redox solution preparation}

We choose ferrocene dimethanol (~\ch{Fe(MeOH)2}), which has a much higher solubility than~\ch{FeMeOH}, in fact, the water solubility of~\ch{Fe(MeOH)2} is around 100~mM~\cite{petrovic_determination_2002, kasper_important_2017}. The reason is that~\ch{Fe(MeOH)2} has two -OH groups (rather than only one), which makes it more soluble in water than FeMeOH. However, even though ~\ch{Fe(MeOH)2} has a theoretical solubility of 100~mM, it is still not easy to realise concentrations beyond 20 mM in the lab, and moreover the dissolution rate of ~\ch{Fe(MeOH)2} in water is super slow. To make the high ~\ch{Fe(MeOH)2} concentration solution of 10mM for our experiments, we dissolved \ch{Fe(MeOH)2}  in hot water (70 \textdegree C), applied magnetic stirring to speed up the dissolution process, and then slowly cooled down to room temperature.

\subsection*{Nanohole Fabrication}

The process for the fabrication of nanohole electrodes is shown in Fig. S13 in $SI~Appendix$.
First of all, borosilicate glass coverslips (No. 1.5H, $24\times 50$~mm, Paul Marienfeld GmbH) were cleaned by rinsing them sequentially with DI~\ch{H2O}, ethanol, DI~\ch{H2O}, isopropanol, DI~\ch{H2O}, ethanol, DI~\ch{H2O} and drying under a clean stream of ~\ch{N2}.
Then the cleaned coverslips were put in the~\ch{O2} plasma surface cleaning machine (ZEPTO, Diener electronics) for $10$ minutes within a 3~mBar atmospheric environment. 
Secondly, a $100$~nm Au layer was deposited by an Electron Beam Physical Vapor Deposition system (EBPVD, Flextura M508E, Polyteknik) on the cleaned glass coverslips through a homemade shadow mask.
To increase the adhesion between the Au layer and the coverslips, a $5$~nm Ti layer was deposited on the cleaned coverslips before the Au deposition using the EBPVD method.
Next, a $50$~nm thick ~\ch{SiO2} layer was deposited on top of the Au layer using the EBPVD method through a second homemade shadow mask to partially expose the Au layer for electric connection.
Finally, the sample with deposited layers was put in the Dual-beam scanning electron microscopy-focused ion beam (SEM-FIB) instrument (Helios G3 Nanolab, Thermo Fisher) to drill the array of nanoholes.
FIB drilling was performed at $30$~KV~\ch{Ga+} ion acceleration voltage, $24$~pA ion current, $50~\mu$s dwell time, and $100$~nm Z-depth to drill the nanohole array.

\subsection*{PNP-BV equations}
The Poisson-Nernst-Planck (PNP) equations and Butler-Volmer (BV) equations are used to solve the diffusion, migration, and reaction processes self-consistently. The Poisson equation can be expressed as 
\begin{equation}
\nabla \cdot (\varepsilon_0 \varepsilon_r \nabla \phi) = -e \sum_i z_i C_i,
\end{equation}
where $\phi({\bf r},t)$ denotes the electrostatic potential, $C_i({\bf r},t)$ and $z_i$ are the concentration and valency of species $i$, respectively, and $e$ and $\varepsilon_r=78$ denote the elementary charge and the relative dielectric constant of water, respectively. The Nernst-Planck equation for the flux ${\bf J}_i({\bf r},t)$ of species $i$ at fixed room temperature $T$ is written as the sum of a diffusive and migration term,
\begin{equation}
\mathbf{J}_i=-D_i\nabla C_i- \frac{D_i z_i e C_i}{k_B T}\nabla\phi,
\end{equation}
with $D_i$ the diffusion coefficient of species $i$ and $k_B$ the Boltzmann constant. The concentrations and fluxes of all species are coupled by the continuity equation 
\begin{equation}
\frac{\partial C_i}{\partial t} = -\nabla \cdot \mathbf{J}_i,
\end{equation}
where the focus here is on the four species in the solution labeled by
%where $C_i, D_i, z_i$ and $\mathbf{J}_i$ is the concentration, diffusivity, valency and flux, respectively, including the solution species  of~
$i=\,$\ch{K+},~\ch{Cl-},~\ch{Fc} and~\ch{Fc+}. Their valencies are taken to be $z_{\text{\ch{K+}}}=+1$, $z_{\text{\ch{Cl-}}}=-1$, $z_{\text{\ch{Fc}}}=0$, and $z_{\text{\ch{Fc+}}}=+1$, and their diffusion coefficients as $D_{\text{\ch{K+}}}=1.97 \times 10^{-9} \mathrm{m^{2}/s}$, $D_{\text{\ch{Cl-}}}=2.03 \times 10^{-9} \mathrm{m^{2}/s}$, and $D_{\text{\ch{Fc}}}=D_{\text{\ch{Fc+}}}=6.7 \times 10^{-10} \mathrm{m^{2}/s}$ \cite{Zhang047}. With appropriate initial and boundary conditions, including the redox reactions on the electrode-electrolyte interface to be discussed below, this set of PNP equations is closed and can be solved numerically in the geometry of a single hole in contact with the grounded bulk electrolyte as depicted in Fig.\ref{Model}A. 

The concentration-dependent BV equation for the electric current density $j$ is given by
\begin{equation}
j = e k_0 \left[ C_{0, \mathrm{Fc}} e^{\frac{e \alpha \eta}{k_B T}} - {C_{0, \mathrm{Fc}^+}} e^{-\frac{e \alpha \eta }{k_B T}}  \right],
\end{equation}
where the time-dependent over-potential is $\eta(t) = V(t) - \phi_{\text{eq}}(t)$, the electrolyte equilibrium potential is $\phi{_\text{eq}(t)} = E_0 + \frac{k_BT}{e} \ln \left( \frac{C_{0, \mathrm{Fc}^+}(t)}{C_{0, \mathrm{Fc}}(t)} \right)$, $C_{0, \mathrm{Fc}/ \mathrm{Fc}^+}(t)$ is the surface concentration of $\mathrm{Fc}$ and $\mathrm{Fc}^+$, $E_0$ = 50 mV is the standard reduction potential of this redox couple, $\alpha$ = 0.5 and reaction speed  $k_0$ = $10^{-4}$ $\mathrm{m/s}$ \cite{Guin444, Zhen351}. The following boundary conditions are set to solve the PNP-BV equations. At the reaction boundaries (AB, BC, FG and GH, where we denote the potential and the fluxes as $\phi_0$ and ${\bf J}_{0,i}$, respectively), a neutral Stern layer of thickness $\lambda_S=0.4\text{~nm}$ is considered that separates the bulk electrode at potential $V(t)$ from the reaction plane, 
\begin{equation}
\phi_0 + \lambda_\mathrm{S} (\mathbf{n}\cdot \nabla \phi)=V(t), 
\end{equation}
and a flux boundary condition is considered for the consumption and product of reaction species, 
\begin{equation}
\mathbf{J}_{0,\mathrm{Fc}^+}=-\mathbf{J}_{0,\mathrm{Fc}}=j/e.
\end{equation}
At the bulk boundary DE, (with potential $\phi_b$ and concentrations $C_{b,i}$ for species $i$), a zero potential and fixed concentration conditions are considered given by 
\begin{equation}
\phi_b=0; C_{b, \mathrm{Fc}}=C_{\mathrm{Fc}};C_{b, \mathrm{Fc}^+}=0;C_{b, \mathrm{{Cl}^-}}=C_{b, \mathrm{K^+}}=C_{\mathrm{KCl}},
\end{equation}
with $C_{\mathrm{Fc}}$ and $C_{\mathrm{KCl}}$ to be specified.  At all other boundaries (AH, CD and EF), zero flux and zero surface charge conditions are set,
\begin{equation}
\mathbf{J}_i=0; \nabla \phi=0.
\end{equation}

%\bibliography{bibfile}
\bibliography{references}
\end{document}